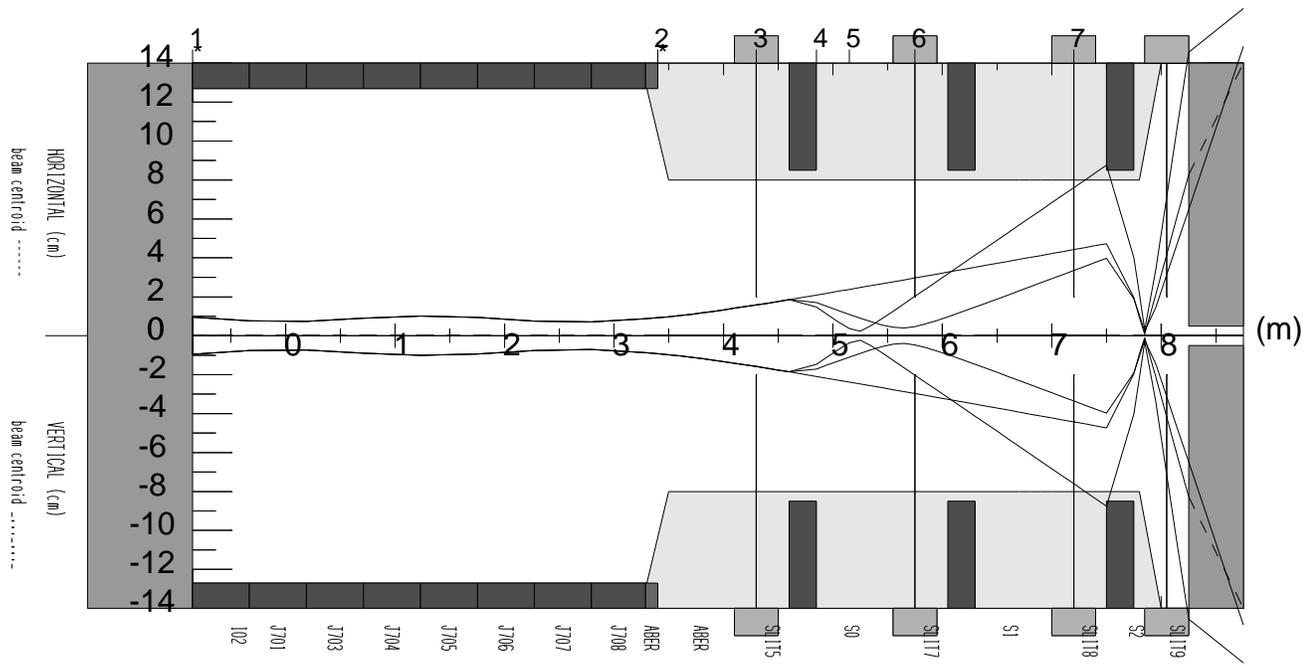

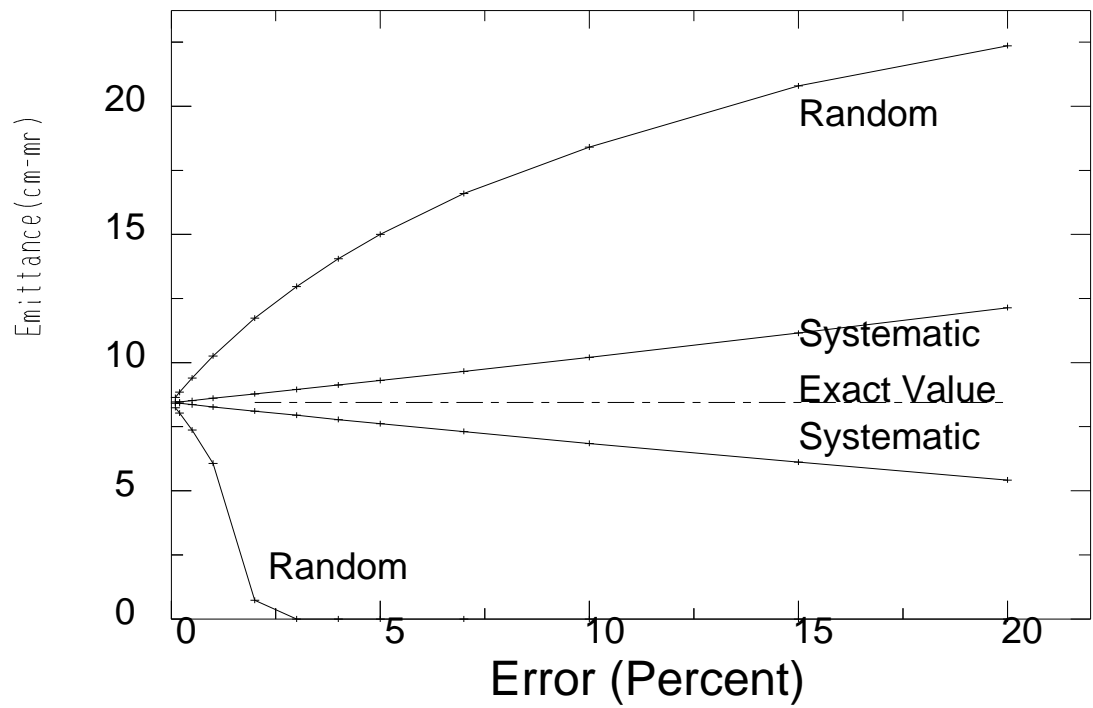

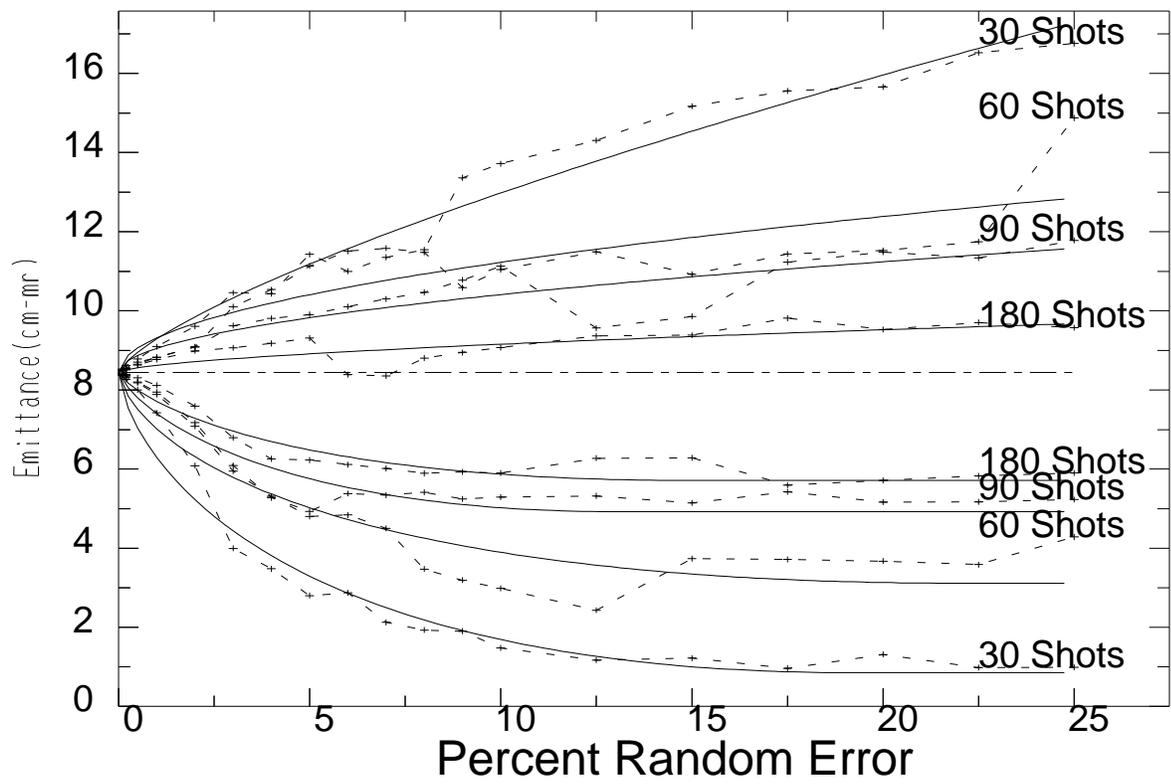

# RECONSTRUCTION OF INITIAL BEAM CONDITIONS AT THE EXIT OF THE DARHT II ACCELERATOR*

Arthur C. Paul, LLNL, Livermore, CA 94550, USA

*Abstract*

We consider a technique to determine the initial beam conditions of the DARHT II accelerator by measuring the beam size under three different magnetic transport settings. This may be time gated to resolve the parameters as a function of time within the 2000 nsec pulse. This technique leads to three equations in three unknowns with solution giving the accelerator exit beam radius, tilt, and emittance. We find that systematic errors cancel and so are not a problem in unfolding the initial beam conditions. Random uncorrelated shot to shot errors can be managed by one of three strategies: 1) make the transport system optically de-magnifying; 2) average over many individual shots; or 3) make the random uncorrelated shot to shot errors sufficiently small. The high power of the DARHT II beam requires that the beam transport system leading to a radius measuring apparatus be optically magnifying. This means that the shot to shot random errors must either be made small (less than about 1%) or that we average each of the three beam radius determinations over many individual shots.

## 1 THE DARHT II BEAMLINE

The DARHT II beamline[1] consists of a series of transport solenoid lens and a kicker system to chop the beam to be sent to the X-ray converter target. Between the accelerator exit and the kicker is a series of three solenoids, lens S0, S2, and S3. Lens S3 matches the beam to the kicker system. Solenoid S2 is used in conjunction with an insertable beam dump, the "shuttle dump", to blow the beam up to a point that the density of energy deposition in the dump is small enough to allow the dumps survival. Solenoid S0, between the accelerator and S2, is used to generate different beam transport conditions for unfolding the initial beam conditions at the exit of the accelerator. A viewing port just in front of S2 is used to measure the beam radius. The beam must be several cm in radius to allow the survival of the viewing foil. The beam exiting this foil has been scattered to the point, that solenoid S2 and the large beam emittance induced by scattering in the viewing foil is sufficient to diverge the beam on the shuttle dump.

## 2 THE PROCEDURE

We define several terms that will be used in this work in order to avoid ambiguity. A "shot" is a 2 usec beam pulse from the accelerator. At minimum, three shots are required to re-construct the initial beam conditions. A measurement of radius can be either a single radius measurement or the average value of many individual measurements. To avoid confusion, we will use the term "determination" to be the measured radius value used in the procedure of beam reconstruction. Three radius determinations are required to re-construct the beam parameters. These three determinations require three shots if each determination is made using a single shot, or 3N shots if each determination is the average of N radius measurements for each determination. The procedure of re-construction unfolds from the determinations the beam emittance, initial radius and tilt. A single unfolding yields a value for the beam emittance, initial radius and tilt. Several unfoldings can be averaged to give an improved value of these parameters.

Consider solenoid S2 set to 7.5 kG to expand the beam onto the shuttle dump almost independent of how the beam address S2. Consider a viewing foil to be inserted into the beamline at the pump port just in front of S2. This is the location were we make the radius measurements. Take the point for beam re-construction to be located 0.1111 meters beyond the exit of the accelerator. The transport of the beam from the re-construction point to the view port is then given by a field free region L1, solenoid lens S0, and field free region L2.

$$V = \begin{bmatrix} 1 & L_2 & 0 & 0 \\ 0 & 1 & 0 & 0 \\ 0 & 0 & 1 & L_2 \\ 0 & 0 & 0 & 1 \end{bmatrix} \begin{bmatrix} C^2 & k^{-1}SC & SC & k^{-1}S^2 \\ -kSC & C^2 & -kS^2 & SC \\ -SC & -k^{-1}S^2 & C^2 & k^{-1}SC \\ kS^2 & -SC & -kSC & C^2 \end{bmatrix} \begin{bmatrix} 1 & L_1 & 0 & 0 \\ 0 & 1 & 0 & 0 \\ 0 & 0 & 1 & L_1 \\ 0 & 0 & 0 & 1 \end{bmatrix} \begin{bmatrix} x \\ x' \\ y \\ y' \end{bmatrix} \quad 2.1$$

$$R = \begin{bmatrix} C^2-kL_2SC & (L_1+L_2)C+k^{-1}S-L_1L_2kS & SC-kL_2S^2 & (k^{-1}-kL_1L_2)S^2+(L_1+L_2)SC \\ -kSC & C-L_1kS & -kS^2 & SC-kL_1S^2 \\ kL_2S^2-SC & (kL_1L_2-k^{-1})S^2-(L_1+L_2)SC & C^2-kL_2SC & (L_1+L_2)C^2+(k^{-1}-kL_1L_2)SC \\ kS^2 & -SC+kL_2S^2 & -kSC & C^2-kL_1SC \end{bmatrix}$$

Here $L_1$, $L_2$ are the drifts between the re-construction point and S0, and S0 and the view port respectively, $C = \cos\theta$, $S = \sin\theta$, $k = B/(2B\rho)$, and $\theta = kL_s$ are the solenoid focusing terms.

Three such transformations are required, one each at a given setting of solenoid S0, say 0, 3.5, and 4.5 kG. Consider three shots with the initial beam conditions of r=0.5 cm, tilt=0, and emittance 3.0 cm-mr, the nominal matched design values. For the three S0 settings above, figure 1, the beam at the view port would be

$$\begin{bmatrix} \sqrt{\sigma_{11}} \\ \sqrt{\sigma_{22}} \\ \sqrt{\sigma_{33}} \\ \sqrt{\sigma_{44}} \end{bmatrix} = \begin{bmatrix} 2.306 \\ 5.792 \\ 2.306 \\ 5.792 \end{bmatrix} \begin{bmatrix} 1.669 \\ 10.350 \\ 1.669 \\ 10.350 \end{bmatrix} \begin{bmatrix} 3.692 \\ 18.705 \\ 3.692 \\ 18.705 \end{bmatrix} \begin{matrix} cm \\ mr \\ cm \\ mr \end{matrix} \qquad 2.3$$

$\sqrt{\sigma_{11}}$ is the horizontal projection of the beam envelope, $\sqrt{\sigma_{33}}$ is the vertical projection of the beam envelope. Here the beam radius is $\sqrt{\sigma_{11}} = \sqrt{\sigma_{33}}$ as the beam is round and in its principle coordinate system, x=y=r.

## 3 THE SIGMA MATRIX

Let $\sigma$ be the matrix characterizing the phase space ellipse bounding all particles in the beam. Let R be the linear transformation matrix from the point of beam reconstruction to the location of the beam size measurement. A point (x,x') on the phase ellipse is given by

$$\sigma_{22} x^2 - 2\sigma_{12} xx' + \sigma_{11} x'^2 = \det(\sigma) \qquad 3.1$$

$$(x, x') \begin{bmatrix} \sigma_{11} & \sigma_{21} \\ \sigma_{21} & \sigma_{22} \end{bmatrix}^{-1} \begin{bmatrix} x \\ x' \end{bmatrix} = 1 \qquad 3.2$$

In the principle coordinate system of a round beam represented by the vector space $(x, x', y, y')^T$ we have $x = y$ and $x' = y'$. If the beam is un-correlated between x and y, then $\sigma_{31} = \sigma_{32} = \sigma_{41} = \sigma_{42} = 0$, and from symmetry $\sigma_{13} = \sigma_{23} = \sigma_{14} = \sigma_{24} = 0$, but possibly tilted in the horizontal and vertical phase space, then $\sigma_{21}$ and $\sigma_{43}$ would be non-zero, and eq(3.2) becomes

$$(x, x', y, y') \begin{bmatrix} \sigma_{11} & \sigma_{21} & 0 & 0 \\ \sigma_{21} & \sigma_{22} & 0 & 0 \\ 0 & 0 & \sigma_{33} & \sigma_{43} \\ 0 & 0 & \sigma_{43} & \sigma_{44} \end{bmatrix}^{-1} \begin{bmatrix} x \\ x' \\ y \\ y' \end{bmatrix} = 1 \qquad 3.3$$

The four dimensional linear transformation matrix

$$R = \begin{bmatrix} R_{11} & R_{12} & R_{13} & R_{14} \\ R_{21} & R_{22} & R_{23} & R_{24} \\ R_{31} & R_{32} & R_{33} & R_{34} \\ R_{41} & R_{42} & R_{43} & R_{44} \end{bmatrix} \qquad 3.4$$

transforms the sigma matrix by the similarity transform

$$\sigma = R \, \sigma_o \, R^T \qquad 3.5$$

Define the initial reconstruction sigma matrix elements of $\sigma_o$ for a round beam to be: $a \equiv \sigma_{11_o} = \sigma_{33_o}$, $b \equiv \sigma_{21_o} = \sigma_{43_o}$, $c \equiv \sigma_{22_o} = \sigma_{44_o}$. The square of the measured round beam size $r^2 \equiv \sigma_{11} \equiv \sigma_{33}$ is given by

$$r^2 = \left[ R_{11}^2 + R_{13}^2 \right] a + 2 \left[ R_{11} R_{12} + R_{13} R_{14} \right] b + \left[ R_{12}^2 + R_{14}^2 \right] c \qquad 3.6$$

Here we have explicitly expanded eq(3.5) in terms of matrix eq(3.4) to represent $\sigma_{11}$ in terms of the initial beam sigma elements (a,b,c) and the values of the transformation matrix. Three sets of radius determinations, $r_1$, $r_2$, and $r_3$ allow reconstruction of the initial beam parameters, a, b, and c.

$$r_1^2 = C_{11} a + C_{12} b + C_{13} c$$
$$r_2^2 = C_{21} a + C_{22} b + C_{23} c \qquad 3.7$$
$$r_3^2 = C_{31} a + C_{32} b + C_{33} c$$

$C_{11}, C_{12}, \ldots$ are the known combinations of the transformation matrix elements for the jth determination. $C_{j1} = \left[ R_{11}^2 + R_{13}^2 \right]_j$, $C_{j2} = \left[ R_{11} R_{22} + R_{13} R_{14} \right]_j$, $C_{j3} = \left[ R_{12}^2 + R_{14}^2 \right]_j$

Inverting we have the desired initial reconstructed beam parameters.

$$\begin{bmatrix} \sigma_{11_o} \\ \sigma_{21_o} \\ \sigma_{22_o} \end{bmatrix} = \frac{1}{\det C} \begin{bmatrix} C_{22} C_{33} - C_{23} C_{32} & -C_{12} C_{33} + C_{32} C_{13} & C_{12} C_{23} - C_{13} C_{22} \\ -C_{21} C_{33} + C_{23} C_{31} & C_{11} C_{33} - C_{13} C_{31} & -C_{11} C_{23} + C_{13} C_{21} \\ C_{21} C_{32} - C_{31} C_{22} & -C_{11} C_{32} + C_{31} C_{12} & C_{11} C_{22} - C_{12} C_{21} \end{bmatrix} \begin{bmatrix} r_1^2 \\ r_2^2 \\ r_3^2 \end{bmatrix}$$

## 4 RECONSTRUCTION

Lets consider a case of beam reconstruction using three shots with each radius determination subject to a systematic error. As each radius determination would have the same error, the reconstructed initial beam radius should systematically have that same error, and the beam emittance being an area should have twice that error. Define the systematic measured radius error to be $\delta r$, then

$$\frac{\delta \varepsilon}{\varepsilon_o} = 2 \frac{\delta r}{R} \qquad 4.1$$

Lets now consider a case of un-correlated random errors in the three shots used to reconstruct the beam initial conditions. If the beam size is magnified by the "optics" of the lens system, then a measurement error $\delta r$ will be magnified by the system magnification M. As the three measurements are un-correlated, the magnified errors will not cancel and we expect, for small errors that the emittance error should grow as

$$\frac{\delta \varepsilon}{\varepsilon_o} = 2M \frac{\delta r}{R} \qquad 4.2$$

Note for small $M\delta r$, that the emittance curve is a straight line with slope M times that of the systematic error curve, figure 2. For large $M\delta r$ the emittance curve parallels the systematic error curve. The transition between these two regimes appears to be some fractional power of the parameter $M\delta r$.

$$\frac{\varepsilon}{\varepsilon_o} = 1 + a z^n - b z^{2n} + \ldots \qquad 4.3$$

$$z \equiv M \frac{\delta r}{R} \qquad 4.4$$

Figure 2 shows the unfolded beam emittance as a function of the magnitude of the un-correlated random error in the beam size determination. The shape of the curve is approximated by eq(4.3) with a=0.2, b=0.004, and n=2/3. Let M be some measure of the optical magnification of the system and $\delta r$ be the radius error.

$$\frac{\varepsilon}{\varepsilon_o} = 1 + 0.2z^{2/3} - 0.004\, z^{4/3} \qquad 4.5$$

Consider our example with M=5.39 $\delta r/R$=10%, $\varepsilon_o = 8.44$, a=0.2, n=2/3, then

$$\frac{\delta\varepsilon}{\varepsilon_o} = anMz^{n-1}\frac{\delta r}{R} \qquad 4.6$$

$$\delta\varepsilon = 1.90\varepsilon_o = 16.0 \text{ cm-mr}$$

almost a 100% error in the reconstructed emittance.

## 5 NUMBER OF REQUIRED SHOTS

We consider two strategies for rendering the beam emittance from N shots. The first uses one unfolding of the emittance based on three radius determinations were each radius determination is the average of N/3 shots. Call this scenario A. The second strategy is based averaging N/3 emittance unfoldings each of which are the result of three radius determination with each radius determination consisting of a single shot. Call this scenario B. The fit to the emittance error curves are represented by eq(5.1)

$$\varepsilon = a_1\, x + a_2\, x^n \qquad 5.1$$

x is the error in percent and $\varepsilon$ is the emittance in cm-mr. This equation is used to fit scenarios A and B. Scenario A is well represented by a 0.72 power law. Scenario B is represented by a 0.5 - 0.6 power law. Note that the lower bounds, are straight lines for large errors given a lower bound to the emittance independent of the value of the random error, figure 3.

The reason that making many emittance determinations with out averaging the radius values gives an better value for the average value of the unfolded beam emittance is that values of emittance that by the luck of the draw (random number sequence) that are negative or zero are averaged out by the positive random values. In scenario A, were we unfold just one emittance but average the radius determinations yield a negative or zero value for some random sequences. With just one unfolding there are no positive values to average this unfortunate value.

## 6 CONCLUSIONS

The beam emittance is related to the area in phase space occupied by the particles comprising the beam. The reconstruction of the emittance by a radius measurement with error $\delta r$ should yield an error in the emittance of at most $2\delta r$. With a system of optical magnification M the error is $2M\delta r$. Use of the shuttle dump diagnostic on DARHT II to determine the beam emittance to within a factor of two using a minimum number of shots requires either 1) the random uncorrelated shot to shot errors be less than about 1%, or 2) we average 30 to 60 shots using scenario B with random errors some where in the range of 5 to 10 percent.

Figure 1. Over plot of the three beam envelopes for the example of beam reconstruction with solenoid S0 at 0, 3.5, and 4.5 kG. The radius measurement is made at view port located at 7.20 meters.

Figure 2. Unfolded beam emittance vs random error in the beam radius measuremet. The exact value and the range of values for a systematic error is also shown. The radial magnification of 5.4 amplifies the error.

Figure 3. Required number of shots for a given emittance range vs radial error using averaging of emittance value method. Nominal beam radius 0.95 cm, tilt -0.35, and emittance of 8.44 cm-mr.